\documentclass[12pt]{article}
 
\usepackage{epsfig}
\begin{document}

\newcommand{\be}{\begin{equation}}
\newcommand{\ee}{\end{equation}}
\newcommand{\bea}{\begin{eqnarray}}
\newcommand{\eea}{\end{eqnarray}}
\newcommand{\da}{\dagger}
\newcommand{\dg}[1]{\mbox{${#1}^{\dagger}$}}
\newcommand{\hlf}{\mbox{$1\over2$}}
\newcommand{\lfrac}[2]{\mbox{${#1}\over{#2}$}}
\newcommand{\scsz}[1]{\mbox{\scriptsize ${#1}$}}
\newcommand{\tsz}[1]{\mbox{\tiny ${#1}$}}


\begin{center}

\large{\bf Measuring the Interplanetary Medium with a Solar Sail}

\vspace{0.5in}

\normalsize
\bigskip 

Michael Martin Nieto${^1}$, Slava G. Turyshev$^2$\\

\normalsize
\vskip 15pt

${^1}$Theoretical Division (MS-B285), Los Alamos National Laboratory,\\
University of California,  Los Alamos, New Mexico 87545, U.S.A. \\
E-mail: mmn@lanl.gov         

\vskip 5pt
$^{2}$Jet Propulsion Laboratory, California Institute of  Technology,\\
Pasadena, CA 91109, U.S.A. \\ 
Email: turyshev@jpl.nasa.gov

\end{center}

\baselineskip=.33in

\begin{abstract}

NASA has been considering a solar sail that would accelerate a craft to
a high velocity ($\sim$14 AU/yr) by the time it reached 5 AU. Then the 
sail would be dropped and  
the craft would coast alone to deep space.  We propose that the
sail be retained longer.  Then the density of the interplanetary
medium could be determined by measuring the 
drag force on the huge sail using radiometric navigational data.  
Such an experiment would yield an independent, 
new type of measurement of the interplanetary medium and should be 
pursued.

\end{abstract}

\begin{center}
\today
\end{center}

\newpage

\section{Introduction}

Sending a spacecraft beyond the heliopause to begin the exploration of our
local galactic neighborhood will be one of the grand scientific enterprises
of this century.
NASA's InterStellar Probe \cite{garner2000}-\cite{liewer2002}, 
for example, could be the first spacecraft
that investigates the nearby interstellar medium and its interaction with
our solar system.  We point out here that the sail could also
be used for an independent determination of the interplanetary medium by
measuring the drag force on it as it passed through the solar system.

The space between the planets in our solar system is far from empty.  
We know that the interplanetary medium has the following
structure: 
First there is the electromagnetic part, composed of the 
(i)  solar radiation (photons) and 
(ii) the magnetic field (also primarily originating at the Sun).
The material components of the interplanetary medium consist of
thinly scattered matter in the form of 
(iii) neutral hydrogen and microscopic dust particles and (iv) a hot 
plasma of electrically charged particles (chiefly protons and
electrons produced by ionization), a.k.a. the solar  wind.  

While the Sun's radiation is obvious,
the other components of the interplanetary medium were not directly
discovered until the era of modern space exploration.  
The exact composition of the material content is still
being debated and many models as to what what exactly is its  nature and 
origin have been proposed.  Whatever the total picture, it is the  Kuiper belt
that is the most significant.

In 1951 Kuiper suggested that some comet-like debris from the formation of
the solar system must exist beyond Neptune.  This since-discovered ``Kuiper
Belt'' is a disk-shaped region past the orbit of  Neptune, roughly 30 to 100
AU from the Sun, containing dust and many small icy bodies. It is now
considered to be the source of the short-period comets.  (The long-period
comets are believed to be formed further away in the Oort cloud.)

There currently is great interest in understanding the Kuiper belt region.
In particular, the study of the trans-Neptunian asteroids is a rapidly 
evolving field of research, with major observational and theoretical
advances in the last few years.   Further interest in the Kuiper belt
comes from disk-shaped regions of dust  
having been observed around other stars in several systems. 

The currently envisioned missions to the edge of the solar
system will, of course, travel through the Kuiper belt.  
For a solar sail mission, the sail is first be used to bring the
spacecraft in to 0.25 AU. Then, as 
the craft is rotated to be directly ``upwind'' it is subjected
to a large outward radiation pressure 
that initiates its journey to the interstellar region. 

The standard concept is to jettison the sail at approximately 5 AU, 
when the craft is already near its terminal velocity (see Section
\ref{drag}), thereby avoiding any later complications with the sail. 
The spacecraft then coasts, 
hopefully staying alive until it reaches 200 -- 400 AU.  It would 
explore the Kuiper belt, the boundaries of the heliosphere, and the
nearby interstellar medium. 
Solar sail propulsion was selected for this mission because of 
recent dramatic advances in solar sail materials development. 

However, if the sail were {\it not} jettisoned at 5 AU, but
kept, if would provide an excellent opportunity to ``directly"
measure the density of matter (dust and gas) in the Kuiper belt by 
observing the drag produced on the sail.  Here we will
analyze the  potential application of a solar sail for 
detecting the properties of the interplanetary medium. 

We will examine whether interplanetary matter, such as that  in the
Kuiper belt would produce a measurable drag on a solar sail.  
We find that a solar sail mission might be extremely useful in
directly determining the amount of dust and gas in the deep solar system. 
Indeed, we show that a modest navigation effort (much less than the
state of the art) will determine the  orbital 
parameters of the sail to the needed accuracy.  
Therefore, this novel space travel technology would 
offer a unique instrument to study the dust and particle content
in the distant regions of our solar system, 
at little additional cost to a planned primary mission.


\section{Major Forces on the Sail}
\label{majorforces} 

For definiteness in our analysis we will consider 
a project with the characteristics of the InterStellar Probe Mission 
\cite{garner2000}-\cite{liewer2002}.  
In the mission concept developed by NASA's
InterStellar Probe Science and Technology Definition Team, a 400-m diameter
solar sail accelerates the spacecraft to $\sim$ 14 AU/year 
(1 AU/yr = 4.74 km/s).  It has a 
total cruising mass of $m \sim 300$ kg, the 
radius of the sail is $\sim 200$ m, and the radius of the inside craft
mount $\sim$ 5.5 m.  As noted above,  
the sail is nominally to be abandoned at 5 AU. when the craft has a
velocity near its hyperbolic terminal 
velocity of $v_{s}(\infty)\sim 14.1$ AU/year.\footnote{There are also other 
interesting concepts, such as the 
ESA/German Odissee concept \cite{odissee}.}

InterStellar Probe's unique voyage from Earth to beyond 200 AU will enable
the first comprehensive measurements of the plasma, neutrals, dust, magnetic
fields, energetic particles, cosmic rays, and infrared emission in the
outer solar system out to the boundary of the heliosphere and beyond into
the interstellar medium. 
This will allow the mission to address key questions about the
distribution of matter in the outer solar system, the processes by which the
Sun interacts with the galaxy, and the nature and properties of the nearby
galactic medium. 

However, if, as proposed in the introduction, the solar sail was
retained through the inner regions of our solar system, 
such a mission could  also yield very important results on
the dust and particle content on the way to interstellar space.  

We now consider this
particular trajectory phase of the mission, when the craft is still within
the solar system.
The major forces on the craft and sail in this period are gravity and solar
radiation  pressure.   (There is also a very small force due to the
solar wind.) Below we shall discuss the effect of these forces in
detail.  In the next section we will discuss the drag force, whose 
measurement is what we are proposing.  


\paragraph{Gravitational acceleration:}
The Newtonian gravitation acceleration is simple to compute and it is
given by 
\be
a_{\tt G} = -\frac{G~M_\odot}{r^2} = -0.594 
\mathrm{\left[\frac{1~AU}{\it r}\right]^2~cm/s^2}. \label{gac}
\ee
The effects of Jupiter and the other planets,
as well as general relativistic effects, are included in all standard 
Orbital Determination Programs. We ignore them here since we are 
concentrating on the relative sizes of non-inertial forces with respect
to gravitational forces and to navigational precision. 


\paragraph{Solar radiation pressure:}
The acceleration due to solar radiation pressure is also well-known and is
given by 
\begin{equation}
a_{\tt sr} =\mathcal{K}_{\tt sr}~
\frac{f_\odot~ A }{c~m~~r^2},
 \label{eq:srp}
\end{equation}
where  $\mathcal{K}_{\tt sr}$ is the
effective reflection/absorption/transmission coefficient of the sail, 
$f_\odot=1367 ~{\rm W/m}^{2}$(AU)$^2$ is the ``solar radiation
constant'' at 1 AU from the Sun, $A$ is the effective 
area of the craft as seen by the Sun, $c$ is the speed of light, and
$m$ is the mass of the craft.  Designs
call for most of the solar radiation being reflected and the rest
transmitted \cite{garner2000}.
Therefore, given the properties of the proposed
sail materials, $\mathcal{K}_{\tt sr}$ might be in the neighborhood of
1.6-1.8, but we leave it as a parameter.  (Total reflection would have
a  $\mathcal{K}_{\tt sr}= 2$, total absorption would yield $1$, and
total transmission would give $0$.) 

Putting in the numbers we have
\begin{equation}
a_{\tt sr}= 0.191~\mathcal{K}_{\tt sr}~
\mathrm{\left[\frac{1~{\rm AU}}{\it r}\right]^2~cm/s^2}. 
\label{srac}
\ee
So, the solar radiation acceleration is about one-third that of the
Sun's gravity, but in the opposite direction.  


\paragraph{Solar wind acceleration:}
The proton number density of the interplanetary plasma 
is about 5 protons/cm$^3 \rightarrow 0.8 \times 10^{-23}$ g/cm$^3$ 
near the Earth.  This density 
decreases roughly as an inverse-square law farther from the Sun. 
However, the density is highly variable, it 
can be as much as 100 protons/cm$^3$. 
Though very tenuous, it's 
properties can be measured by spacecraft.  Near 1 AU 
the temperature of the interplanetary plasma is about 100,000 K, and
its current velocity $v_{\tt sw}$ is on the order of 
400 km/s.\footnote{The current ``space weather'' near the Earth
can be found at \cite{weather}.}

The 
acceleration caused by the solar wind has the same mathematical form as Eq. 
(\ref{eq:srp}), with $f_\odot/c$ replaced by $m_pv_{\tt sw}^2n_p$, 
where $m_p$ is the proton mass and $n_p \approx5$  cm$^{-3}$ is the
proton number density at 1 AU.
Thus, we have   
\begin{eqnarray}
a_{\tt sw}(r)
&=&\mathcal{K}_{\tt sw}\frac{m_pv_{\tt sw}^2\,n_p\,A} {m \,r^2} 
    \nonumber \\[11pt]
&\approx & 5.61 \times 10^{-5}~ \mathcal{K}_{\tt sw}~
\left[\frac{1~{\rm AU}}{r}\right]^2~{\rm cm/s}^2,
\end{eqnarray}
where again $\mathcal{K}_{\tt sw}$ is an effective 
reflection/absorption/transmission coefficient for the sail. 
Therefore, the effect of the solar wind is much smaller than the  
effects of solar gravity or the solar radiation pressure.  


\section{Drag Forces on the Sail}
\label{drag}

Since any drag force depends on the relative velocity of the craft, 
let us first will review how the velocity of the craft will evolve in
its mission. 

\paragraph{The velocity of the craft in deep space:} 
For a spacecraft in deep space, 
one can compute its velocity to first order using 
standard Newtonian mechanics.  Given that $v_{s}(\infty)= 14.1$ AU/yr, 
the velocity between sail jettison time (foreseen to be at 5 AU) 
and infinity would approximately be (ignoring the angular
momentum energy)
\bea
v_s(r) &\approx&  \left[v_s^2(\infty) + \frac{GM_\odot}{2r}\right]^{1/2}
       \nonumber \\
&\approx& 
\left[14.1~+~\frac{0.7~\mathrm{AU}}{r}
          ~+~ \dots \right]~\mathrm{AU/yr},
\eea
meaning, specifically, that 
\be
v_s(r=5~\mathrm{AU}) = 14.24~ \mathrm{AU/yr}
  \approx 67.5~ \mathrm{km/s}. \label{v5}
\ee

This shows that after 5 AU, the slowing effect of gravity 
would not significantly affect the time of travel to deep
space.  Over time, gravity will decrease the hyperbolic velocity by
only 0.14 AU/yr.  If, on the other hand, the sail remained attached to
the craft after 5 AU, comparing Eqs. (\ref{gac}) and (\ref{srac}) shows
that the 
solar sail would contribute only about +0.05 AU/yr in terminal velocity. 
Therefore, to avoid possible measurement problems the sail ordinarily 
the sail would be abandoned at a distance of about 5 AU from the Sun.  

\paragraph{The drag acceleration at 5 AU and beyond:}
Consider the drag force if the sail remained attached at and beyond 5 AU.  
The acceleration due to drag from the interplanetary medium is 
\be
a_{\tt d}(r)= -\mathcal{K}_{\tt d}\frac{\rho~~v_{s}^2(r)\,A}{m}, 
\label{dac}
\ee
where $\rho(r)$ is the density of the interplanetary medium, 
$\mathcal{K}_{\tt d}$ is the effective
reflection/absorption/transmission coefficient of the sail for the 
particles being hit by the sail, and $v_{s}^2(r)$ is the effective
relative velocity of the craft with respect to the medium.  
The critical unknowns are $\rho(r)$ and $\mathcal{K}_{\tt d}$.

Current limits on the gas and dust in deep-space interplanetary medium
are not precise \cite{teplitz1999,scherer2000}.
But the amount of gas is much less that the amount of dust.\footnote{
The gas is believed to come mainly from the interstellar medium as the
Sun revolves around the galaxy \cite{scherer2000}. 
Therefore, the drag of the gas is
``unidirectional'' in the sense that it has a velocity relative to the
solar system of about 25 km/s.  (Thus, the actual gas 
drag velocity on a spacecraft is the vector sum of the craft's velocity, 
$\approx$ 70 km/s,  and this 25 km/s.) Its
constant density is roughly equal to that of the solar wind at 20 AU,
so about one hydrogen atom per 100 cm$^3$.}   
The amount of dust is believed to be relatively high.  
Further,  most of the dust should be in orbit about the Sun, so the 
drag velocity will effectively be the radial velocity of the craft. 

The final result will also depend on $\mathcal{K}_{\tt d}$, 
which means it depends on the sizes of
the particles and especially on the properties of the sail.  
A determination of what should be used for  $\mathcal{K}_{\tt d}\rho(r)$
is in practice very difficult.  There is a complicated distribution
of dust grains of various sizes.  Symbolically, one can write 
\be
\mathcal{K}_{\tt d}\rho(r) 
= \int_0^\infty \mathcal{K}_{\tt d}(m)~\rho(r,m)~dm.
\ee
For now we just leave it as a constant. 

Therefore, with the spacecraft properties we are using, and taking the 
value for $v_{s}(r)$ to be  67.5 km/s, as given by Eq. (\ref{v5}),
we find 
\be
a_{\tt d}(r)= - 1.91 \times 10^{-3}~ \mathcal{K}_{\tt d}~
\left[\frac{\rho(r)}{10^{-20}~{\rm g/cm}^3}\right]
~{\rm cm/s}^2.    \label{dacN}
\ee

Since a solar sail would be rotationally stabilized, there would be
little use of attitude-control jetting.  In fact, the sails would be
``trimmed'' (pulled in and out) to affect attitude.  Therefore, 
the precision of the 
navigation could approach that of a true spinner.  This means that with
Doppler and range navigation techniques, as well as the occasional use
of Very Long Baseline Interferometry (VLBI) with the 
differenced Doppler ($\Delta$DOR) technique, modern yet off the shelf
X-band tracking could be precise to less than $10^{-9}$ cm/s$^2$ over a
year \cite{nieto2003}.  

Further, the forces of 
Sec. \ref{majorforces} all vary as 
$1/r^2$.  Given an almost constant velocity, the drag force varies
roughly as the density of the medium.  This makes it even easier to
distinguish a drag force.  

Therefore, even with possible heat systematics , 
with proper craft design  \cite{nieto2003} a density of 1 hydrogen atom
cm$^{-3}$ should easily be seen.  
A measurement of the drag force would provide
information on what might be the content of  
$\rho(r,m)$  \cite{teplitz1999}. 
Given that test experiments on the sail material are done to determine what
$\mathcal{K}_{\tt d}$ is for the interplanetary medium, 
such a measurement could be an independent confirmation to the 
results from any particular ion, gas, and dust detectors on the craft.

This measurement could be done for  $r$ both less than and also greater than
$5~\mathrm{AU}$ (if the sail were retained  for a longer period).  
If there were no overriding reason to get rid of the
sail at 5 AU, we propose it be kept until at least past
30 AU, to reach the Kuiper belt.\footnote{The 
Southwest Research Institute in Boulder maintains 
a bibliography of articles on the Kuiper belt \cite{kuiper}.}  
which would only be about 2 years more of mission time. The amount of new
information obtained with each year would be very valuable.

After the sail is abandoned, the mission would be free to perform its
primary objectives unencumbered, be they those of the InterStellar Probe 
mission \cite{garner2000}-\cite{liewer2002} 
and/or any other objective.\footnote{An example would be a test 
\cite{nieto2003,anderson2002b} of the Pioneer anomaly
\cite{anderson1998,anderson2002a}.} 


\section{Discussion}

Our Solar system is the end product of the common astrophysical
process of stellar system formation from a protoplanetary disk nebula.
Collisions play a central role in the formation and evolution of planetary
systems, either increasing or eroding the mass of the bodies. The present 
interplanetary dust in the solar system is a result of such 
collisional processes. A deep space mission will
provide an opportunity for both in situ and remote sensing (via infixed
emission) of both interplanetary and interstellar dust in the heliosphere
and in the interstellar medium. 

A deep space mission can also 
determine the mass, composition, and orbital
distributions of dust in the outer solar system,to aid in the 
study of its creation and
destruction mechanisms. It can also search for dust structures associated
with planets, asteroids, comets, and the Kuiper belt.

These studies will
constrain theories of collisional dynamics in the solar system and help
us understand the origin and nature of our solar system,
not to mention other planetary systems.  
Finally, a deep space mission can uniquely give insight into the fundamental
question of the radial extent of the primordial solar nebula and, more
precisely, the extent of the primordial planetesimal disk.  

Some time ago,  Boss and Peale had derived a model for a
non-uniform density distribution in the form of an infinitesimally thin
disc extending from 30 AU to 100 AU in the ecliptic plane
\cite{boss1976}.  More recent infrared observations have ruled 
out more than 0.3 Earth mass of  Kuiper Belt dust in the trans-Neptunian
region \cite{teplitz1999,backman1995,stern1996}.

More recently, two primary mass concentrations at 
39.4 AU and 47.8 AU, corresponding to Neptune upon Pluto 
resonances of 3:2 and 2:1 \cite{malhotra1995,malhotra1996} 
were discovered.  
For a different reason, to obtain a limit on gravitational forces,
three specific mass distributions were also studied including this 
distribution \cite{anderson2002a}, namely:  i) a
uniform distribution, ii) a 2:1 resonance distribution with a peak at 47.8
AU, and  iii) a 3:2 resonance distribution with a peak at 39.4 
AU.\footnote{A
total mass of one Earth mass was assumed, which is significantly larger than
the standard estimates given above.}   

One can combine the results of Refs. \cite{teplitz1999,anderson2002a}
to determine an upper limit on the drag acceleration acting on a 
sail by Kuiper belt dust.  The drag acceleration would be on the order of
$10^{-5}$ cm/s$^2$.  This and the other effects given above 
could easily be observed, 
if the sail were retained into quite deep space, 

Further, these accelerations would not be constant across the 
data range.   Rather, they would show an increasing effect as the 
spacecraft approached into the belt and a decreasing effect as it
receded from the belt, even with a uniform density model in the belt.  
This behavior makes the detection of the Kuiper belt dust contents
with a solar sail a reasonably easy task, which justifies the use of
the solar sail for this objective.

We emphasize that {\it if} a deep-space solar-sail mission flies then, with
modest forethought and planning, the illuminative additional information on
the matter content of the solar disk could easily be obtained.  This
course of action must be seriously considered.  


\section{Acknowledgements}

M.M.N. acknowledges support  by the U.S. DOE.  The work of S.G.T was
performed at the 
Jet Propulsion Laboratory, California Institute of Technology, under
contract with the  National Aeronautics and Space Administration.  



\end{document}